\documentstyle{jaa}
\include{graphicx}
\begin{document}
\title[High Resolution Observations and AO]{High Resolution Observations using 
Adaptive Optics: Achievements and Future Needs} 
\author[K. Sankarasubramanian \& T. Rimmele]%
       {K. Sankarasubramanian$^{1}$ \& T. Rimmele$^{2}$ \\ 
        $^{1}$ISRO Satellite Centre, Bangalore 560 017, India, email: sankark@isac.gov.in \\
       $^{2}$National Solar Observatory\thanks{NSO is operated by the
Association of Universities for Research in Astronomy (AURA) under cooperative
agreement with National Science Foundation (NSF)}, Sunspot, NM - 88349, USA}
\maketitle
\begin{abstract}
Over the last few years, several interesting observations were obtained with the
help of solar Adaptive Optics (AO). In this paper, few observations made using
the solar AO are enlightened and briefly discussed. A list of disadvantages with
the current AO system are presented. With telescopes larger than 1.5m are
expected during the next decade, there is a need to develop the existing AO
technologies for large aperture telescopes. Some aspects of this development are
highlighted. Finally, the recent AO developments in India are also presented.
\end{abstract}

\begin{keywords}
Small-scale structures -- Adaptive Optics -- Multi-Conjugate Adaptive Optics --
High Spatial Resolution
\end{keywords}
\section{Introduction}
Sun provides an unique opportunity for greater understanding of physical
processes happening in stellar atmospheres. The solar atmosphere is structured
to very small scales which are dynamic in nature. Understanding the nature of
these small-scale structures and dynamics may provide important clues to some
puzzling and unanswered questions. It is now very well realised that the
small-scale dynamics can influence the large-scale structures (DeRosa, 2005). In
hydrodynamic condition, two important scales determine the structuring of the
solar atmosphere: (i) Pressure scale height, and (ii) Photon mean free path.
Both these scales correspond to a value of 70km (i.e., about 0."1) at the
photosphere. However, smaller magnetic structures are seen in realistic
numerical magneto-hydrodynamic simulations (Stein \& Nordlund, 2006; Schussler
\& Vogler, 2006). Grid sizes used in these simulations are smaller than 0."1 and
future simulations are aimed to achieve a grid size of 0."02 or better.

Few years before, resolving structures on the Sun to better than an arcsecond
for longer durations were not possible even at good sites due to the Earth's
atmosphere. The scenario changed with the advancement of the Adaptive Optics
(AO) technology. Hence, modern telescopes thrive on improving the spatial,
spectral, and temporal resolutions using an AO system. It may be stated that
most of the recent large solar ground based telescope (aperture $>=$ 50cm)
designs have an integrated AO system. The realisation of the day-time AO
technology lagged behind the night-time due to the difficulties caused by the
extended nature of the source as well as the poor seeing conditions prevail
during the day-time (Rimmele, 2004; Keller, 2005).

The first solar AO system (Acton, 1992 \& 1993) was developed at Lockheed and
tested at the Dunn Solar Telescope (DST). This system worked with high contrast
features like solar pores and hence severely limited the scientific use. The
development of the correlating Shack-Hartmann wavefront sensor, for the
low-order AO system at DST (Rimmele, 2000),  has changed the perspective of the
solar AO. All solar AO systems currently in operation are based on the
correlating Shack-Hartmann sensor. Following Keller (2005), Table~I lists the
successful AO systems at different solar observatories around the globe.
Figure~1 shows an example Sunspot image observed under good seeing condition
using a high-order AO (HOAO) system operated at DST. A small portions of the
bottom penumbral region is blown-up to show the high spatial resolution achieved
with this observation.

\begin{figure}[t]
\hspace*{0.2in}
\includegraphics[scale=0.22]{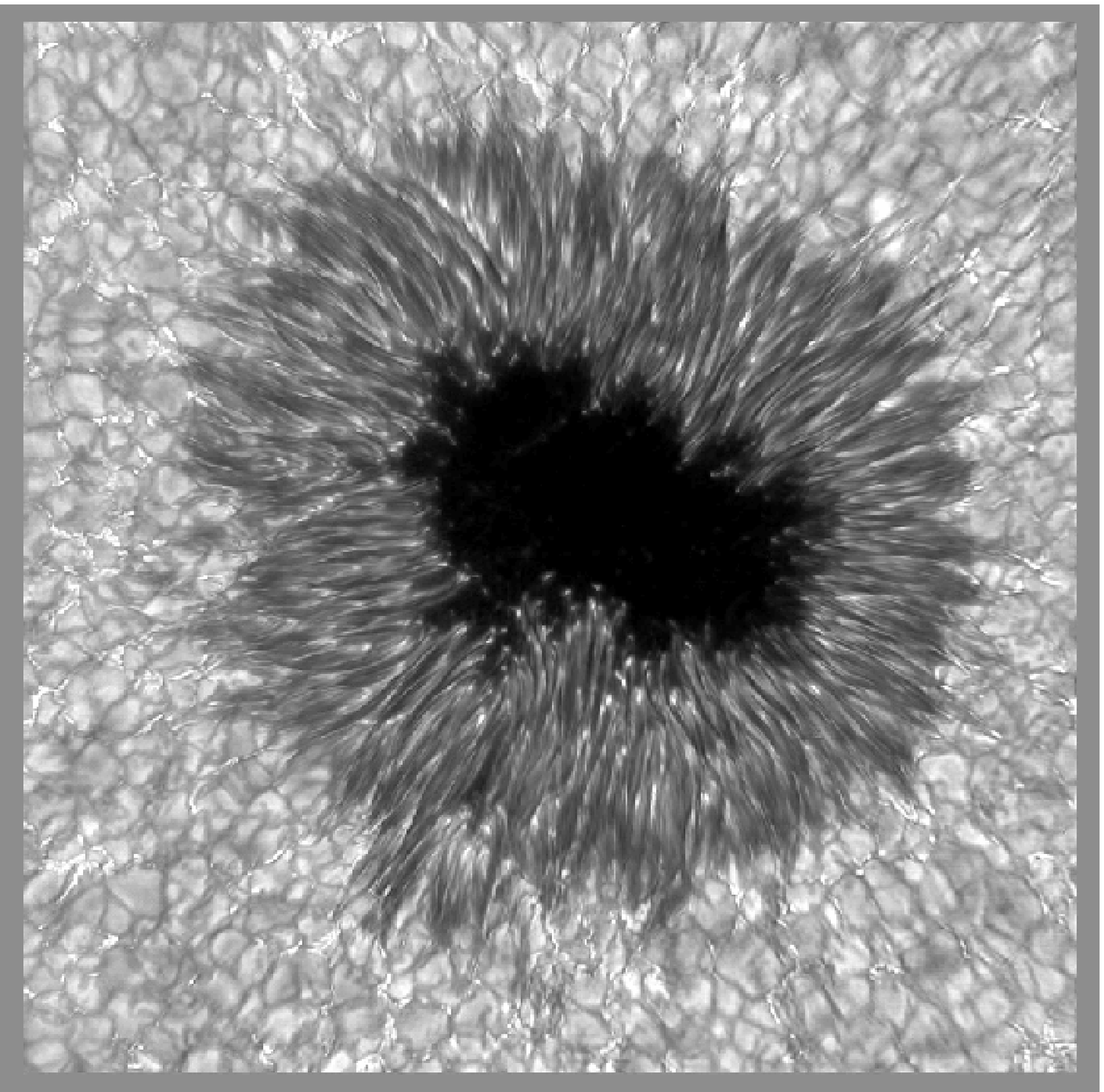}
\includegraphics[scale=0.22]{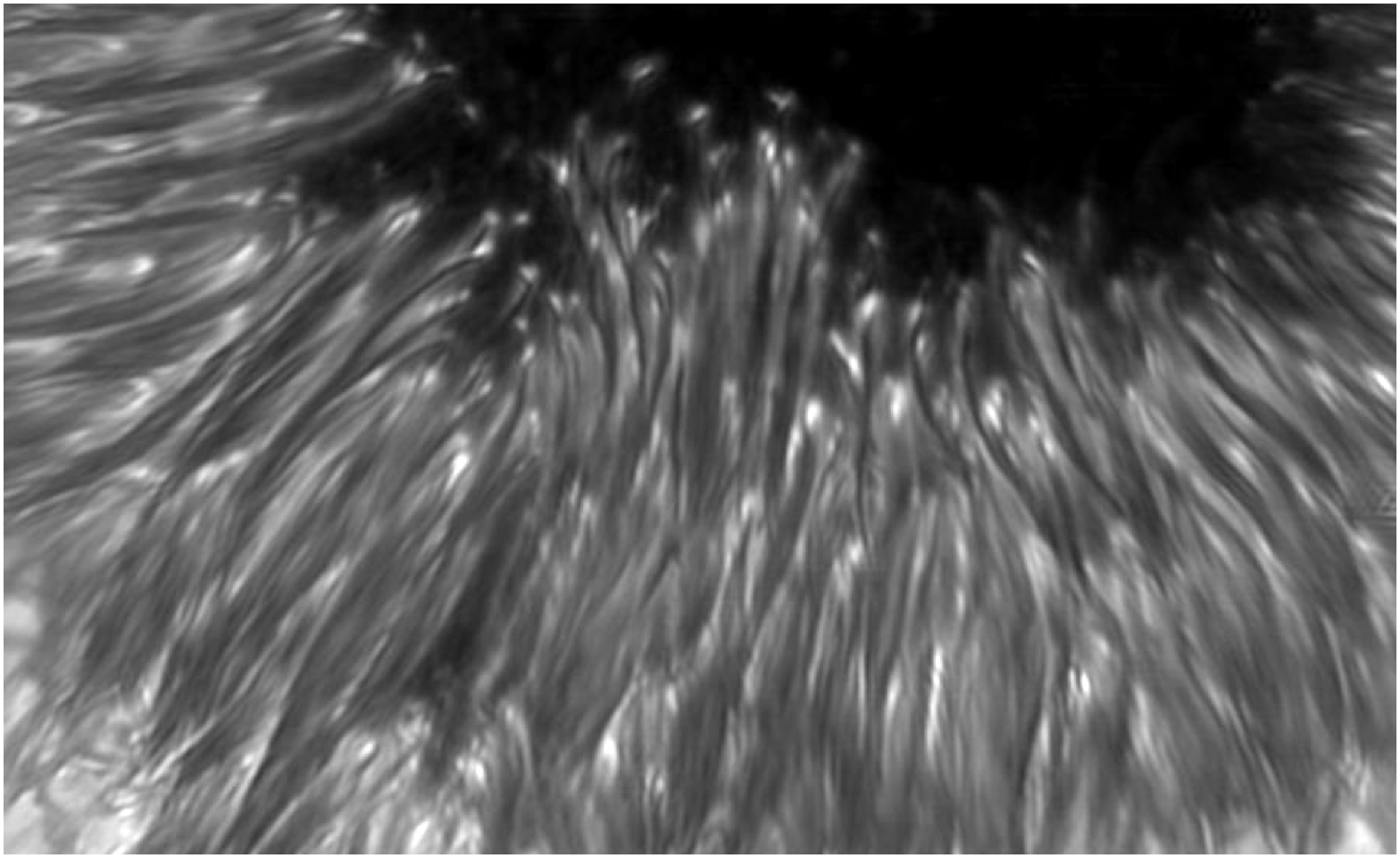}
\caption{Speckle reconstructed sunspot observed at the DST using the high-order
AO (HOAO) system. Courtesy: F. Woeger, T. Rimmele, M. Komsa, C. Berst,
S. Fletcher, \& S. Hegwer: NSO/AURA/NSF.}
\end{figure}

\begin{table}[h]
{\bf Table 1: A list of Successful AO Systems.} \\
{\tiny
\begin{tabular}{|ccccccccc|}
\hline
& & & & & & & & \\
AO & Act- & Sub- & Camera & Frame & Hardware & DM & Ref. & First  \\
& uator & aperture & & Rate & & & & Light \\
& & & & & & & & \\
\hline
& & & & & & & & \\
76cm & 57 & 19 &  ---   & Analog & --- &  ---    & Acton & 1986 \\
DST & & & & & & & 1992 & \\
Lockheed & & & & & & & & \\
& & & & & & & & \\
76cm&  97 & 24 &---& $<$1.6kHz & 24 & Xenitics & Rimmele & 1998 \\
DST & & & & & DSPs &Inc. & 2000 & \\
LOAO & & & & & & & & \\
& & & & & & & & \\
48cm  &  19 & 19 & Dalsa & 955Hz & 566MHz & AOPTIX & Scharmer &  1999 \\
SVST & & & CA-D6& & Alpha &Tech. Inc. & 2000 & \\
& & & & & & & & \\
76cm & 97 & 76 & Custom & 2.5kHz & 40 & Xenitics&Rimmele &	2002 \\
DST & & & built & & DSPs &Inc. &2003 & \\
HOAO & & & CMOS & & & & & \\
& & & & & & & & \\
70cm & 35 & 36 & Dalsa & 955Hz & 8X900MHz & Laplacian & von der Luhe & 2002 \\
VTT & & & CA-D6 & & Sun &Optics &2003 & \\
KAOS & & & & & & & & \\
& & & & & & & & \\
1.5m & 37 & 120-200 & Dalsa & 955Hz & 1GHz &	Okotech  & Keller &	2002 \\
McMath & & & CA-D6 & & Pentium & &2003 & \\
Low Cost & & & & & & & & \\
& & & & & & & & \\
97cm & 37 & 37 & Dalsa & 955Hz & 1.45GHz & AOPTIX & Scharmer & 2003 \\
SST & & & CA-D6 & &Athlon &Tech. Inc. &2003 & \\
& & & & & & & & \\
65cm & 97 & 76 & Custom & 2.5kHz & 40 & Xenitics & Denker & 2004 \\
BBSO & & & Built & &DSPs &Inc. &2007 & \\
HOAO & & & CMOS & & & & & \\
& & & & & & & & \\
\hline
\end{tabular}
}
\end{table}

\section{Science Achievements}
There are innumerable science observations that made use of the AO. Only few
observations are briefly discussed in this paper due to page restriction. The
readers are requested to go through the cited references for details.

\subsection{High Resolution Imaging}
Narrow band G-band images are the first images from most of the AO systems. This
is due to the presence of high contrast small-scale features at this wavelength
band. However, watching the AO corrected images in the broad-band video camera,
attached to the AO system, is always a pleasure for observers. Contrast of these
images are usually enhanced using post processing techniques: speckle
reconstruction, phase-diversity, or long-exposure point spread deconvolution
with or without frame selection. Each of these techniques has its own advantages
and disadvantages.

One of the best high spatial resolution observation achieved during the initial
phase of the AO development is the detection of dark cores in the penumbral
filaments (Scharmer et al., 2002) using the Swedish Solar Telescope (SST). The
observations are obtained using a low order AO system along with real-time frame
selection and subsequent image restoration technique. Observations done with
HOAO at DST confirmed the existence of dark cored penumbral filaments (see
Figure~1). These observations show that the dark cores are unresolved with a
cross-section of about 90km or smaller and their length can be as long as
1000km. They can live for atleast an hour. Apart from the dark cores in the
penumbral filaments, other thin dark structures, named as `canals' and `hairs'
are also observed. There are other interesting observations from the SST over
the last few years. Few examples are: The 3D structures of small-scale fields
(Lites et al., 2004), leakage of photospheric oscillations and flows to form
spicules (de Pontieu, Erdelyi, \& James, 2004), multi-line spectroscopic study
of the dark cored filaments (Bellot Rubio, Langhans, \& Schlichenmaier, 2005),
and the detailed study of the inclination of magnetic fields and flows in the
sunspot penumbrae (Langhans et al., 2005).

\begin{figure}[t]
\begin{center}
\includegraphics[scale=0.50]{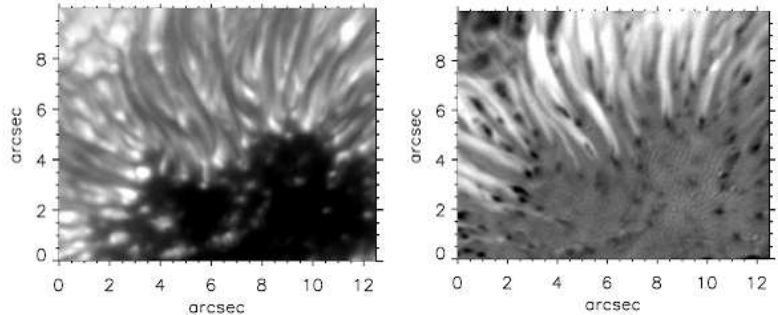}
\caption{Intensity (right) and velocity (left) image of an observed sunspot
penumbral filaments. Dark (bright) in the velocity image represents an upflow
(downflow).}
\end{center}
\end{figure}

High resolution time sequence observation of velocity fields in the penumbral
regions (Rimmele \& Marino, 2006a) has allowed one to quantitatively compare
different penumbral models. The time sequence clearly show that the upflows
(in the footpoints of penumbral dark filaments) and horizontal flow (the
Evershed flow) move around and evolve as a unit, indicating that they are part
of the same feature. These observations also produce strong evidence that the
penumbral grains are the inner footpoints of the Evershed flows where the hot
upflow occurs. A full vector magnetic field observations along with such
velocity observations may well be able to quantitatively differentiate and
refine the few existing models of the penumbra. Such kind of observations can be
expected in the near future. Penumbral asymmetries are also studied in detail
with observations obtained using AO (Tritschler et al., 2004; Soltau et al.,
2005). These recent high spatial resolution penumbral observations have already
started the debate on the suitability of the existing penumbral models (Spruit
\& Scharmer, 2006; Thomas et al., 2006).

Success of the AO has initiated the deployment of several new backend
instruments over the past few years. The Interferometric Bidimensional
Spectrometer (IBIS) is one such instrument deployed at the DST in 2003. This
instrument produces high spatial, spectral, and temporal resolution observations
both in the photosphere and in chromosphere (Cavallini \& Reardon, 2006).
Analysing the time sequence observations obtained with this instrument,
Vecchio et al. (2007) have concluded that waves with frequencies above the
acoustic cut-off propagate to upper layers only in restricted areas of the quiet
Sun. Their results support that the network magnetic elements can channel
low-frequency photospheric oscillations into chromosphere, thus providing a way
to input mechanical energy in the upper layers.

\subsection{High Resolution Spectroscopy}
\begin{figure}[t]
\includegraphics[trim=0 0 50 50,scale=0.24]{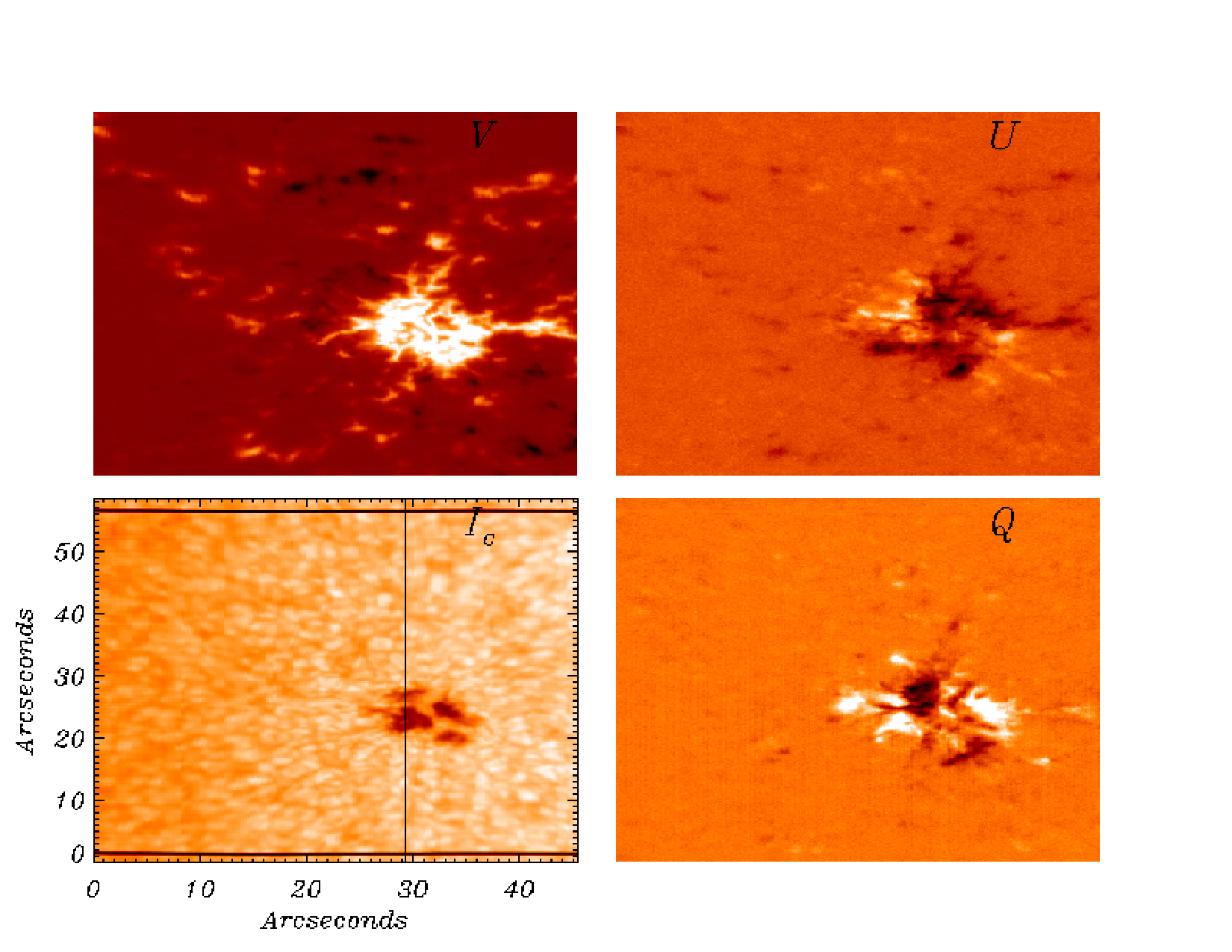}
\includegraphics[trim=0 0 50 50,scale=0.24]{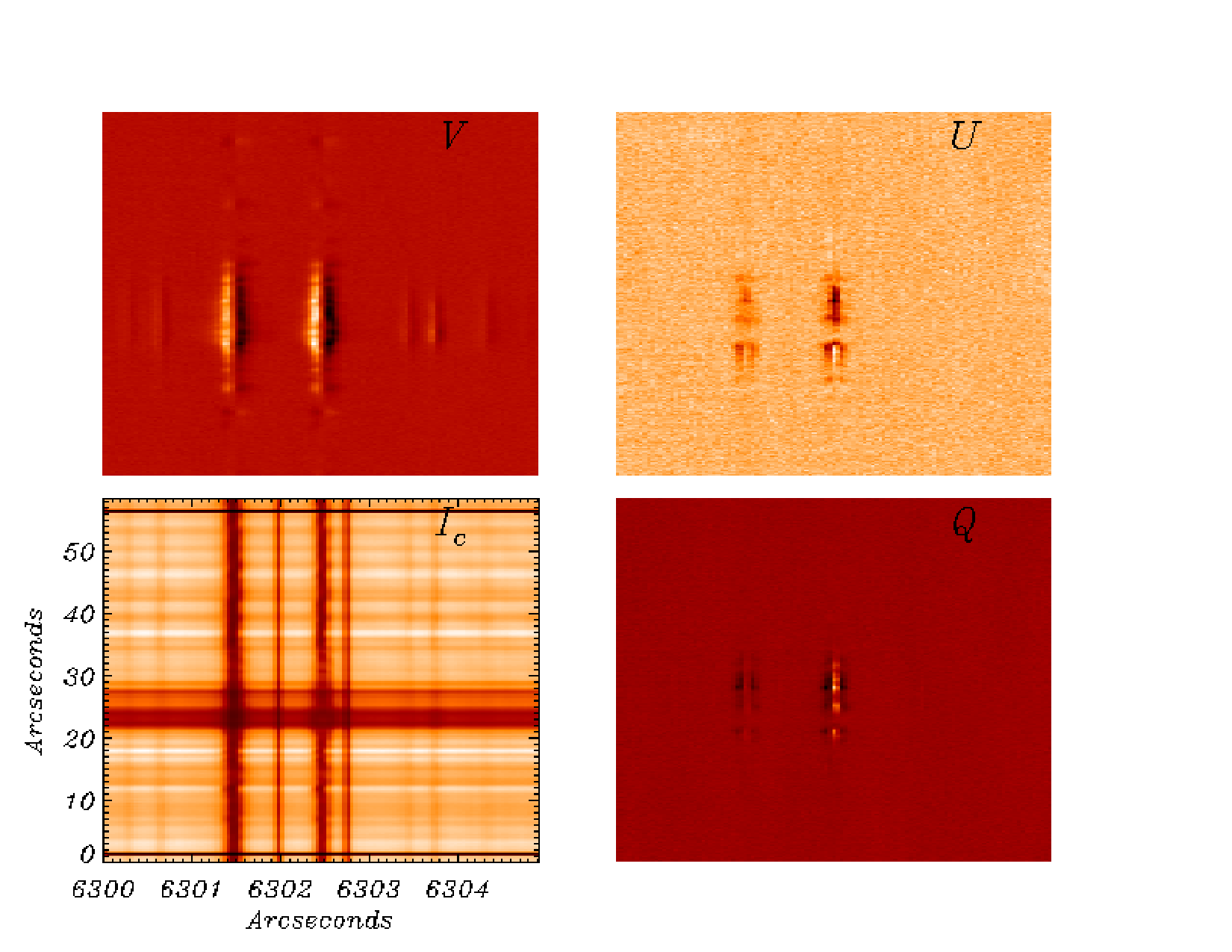}
\caption{Raster image of the continuum intensity and the Stokes images (left).
The maps are produced by averaging a small wavelength region in the continuum
for intensity map and in the blue wing of the Stokes profiles for Q, U, and
V-maps. The right image shows the Stokes spectrum for the slit represented
as a dark vertical line in the continuum intensity image at the left.}
\end{figure}

The real test for any solar AO comes from the spectroscopic and
spectro-polarimetic observations due to the long exposure times and the time
required for scanning the region of interest. Moreover, the raster scan
images obtained with a spectrograph does not undergo any special processing
(like the speckle reconstruction or phase-diversity). In most cases, the solar
AO helps in keeping consistent image quality during the scanning time of the
spectrograph. The scanning time can be as large as 1-hour depending on the
region of interest. The initial observations using the low-order AO corrected
images along with the Advanced Stokes Polarimeter (ASP) has revealed small-scale
needle-like convective structures around a pore (Sankarasubramanian \& Rimmele,
2003). The radius of the ring like upflows seen around the small-pore matched
quiet well with the pattern observed in the magneto-hydrodynamic numerical
simulations (Steiner, 1998). Figure~3 shows a raster image obtained by scanning
the Diffraction Limited Spectro-Polarimeter (DLSP; Sankarasubramanian et al.,
2004) operating at the DST along with the HOAO. This raster image is an example
for the consistent performance of the HOAO for long (45-minute for this case)
duration observations.

Regular observations of diffraction-limited vector magnetic field are feasible
using the HOAO and DLSP. These high spatial resolution vector magnetic field
measurements are started to produce quantitative informations about small-scale
magnetic elements, like umbral dots (Socas-Navarro, 2004; Sankarasubramanian,
Rimmele, \& Lites, 2004), small-scale magnetic elements in and around active
regions (Sankarasubramanian \& Hagenaar, 2007) and quiet sun magnetic fields
(Lites \& Socas-Navarro, 2004). There are other similar spectroscopic
instruments developed at the AO corrected image planes, examples are the POLIS
and TIP (Schmidt et al., 2003; Martinez Pillet et al., 1999). The development of
the AO has also helped in obtaining very high resolution spectroscopic
observations in the near and far infra-red regions (Keller et al., 2003 \& Lin,
private communication).

\section{AO Developments in India}
The development of AO technology was started very recently in India. A low-cost
AO system along with a Shack-Hartmann wavefront analyser was developed and
tested in the laboratory (Ganesan et al., 2005). A 10$\times$10 sub-apertures
for sampling the wavefront and a 37-channel MMDM from OKO Tech was used in this
system. This system cannot be used for solar observations due to the slow 
read-out speed of the camera (25 fps).

\begin{figure}[t]
\begin{center}
\includegraphics[scale=0.50]{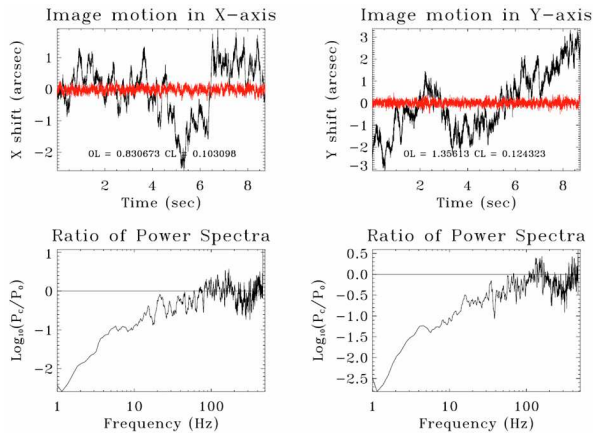}
\caption{A closed-loop tip-tilt correction of the USO-AO system. A closed-loop
update rate of 1kHz is used. Achieved closed-loop bandwidth is about 100Hz and
the improvement in the 'rms' image motion is anywhere between 5 to 25.}
\end{center}
\end{figure}

Udaipur Solar Observatory (USO) is involved in developing an Adaptive optics
system for their 50cm aperture Multi-Application Solar Telescope (MAST;
Venkatakrishnan 2007). A laboratory setup has been developed for testing the
whole system. A 15cm telescope is used as a light feed for the laboratory
optical set-up. A portion of the image from this 15cm telescope is selected, by
using a field stop, for imaging on to a science as well as a wave-front sensing
camera after going through a tip-tilt mirror and a plane mirror (later this
plane mirror will be replaced with a deformable mirror). The wave-front sensing
camera can operate at a rate of 1kHz. The algorithm for estimating the image
motion is implemented using MMX instructions and a PID type control system is
developed for the tip-tilt mirror control. The system was tested in the
laboratory using a laser source as well as using the solar image. Closed loop
tip-tilt correction was achieved and Figure~4 shows the performance of the
system during a closed-loop operation. Experiments are going on to include the
deformable mirror and a closed loop AO correction with the DM is expected to
happen in the middle of 2007.

\section{Future Needs}
The overwhelming scientific observations using the AO effected several solar
observatories (IRSOL, THEMIS, GREGOR, NST, \& USO) to develop an AO system for
their respective telescopes and they are expected to be functional over the next
few years. The two newly proposed large telescope NLST (this proceedings) and
ATST (Wagner et al., 2006) will require a system with an order more compared to
the one currently operating. ATST has proposed for an AO system with 1280
sub-apertures and 1369 actuators and expected to be operated in a closed-loop
bandwidth of 250Hz (Rimmele et al., 2006b).

One of the major limitation of the currently operating AO systems is the
inability to achieve diffraction limited imaging over a larger field-of-view
(FOV). The image becomes blurred away from the isoplanatic patch and the size
of this patch (usually about 10") can vary from site to site and also from day
to day. The difficulty in consistently locking on to the low-contrast features
like granulation is the second major disadvantage with the current AO systems.
The faster evolution of the granules compared to high-contrast structures poses
this difficulty. The performance of the AO reduces when the observations are
done close to the limb and the current AO cannot be used for coronal
observations.

Multi-conjugate Adaptive Optics (MCAO) has been proposed as a technique to
increase the FOV over which the corrections are done. The Sun is an ideal
object to carry out the MCAO due to its extended nature. Solar MCAO efforts
are currently underway at the NSO and at Kipenheuer Institute of Sonnenphysick
(KIS). There are considerable progress made during the last two years in
realising the MCAO (Rimmele et al., 2006c; Berkefeld, Soltau, \& von der Luhe,
2006).

\section{Acknowledgements}
The author KS acknowledges the following personnel for providing the necessary
informations: Balasubramaniam, K. S. (NSO), Bayanna, R. (USO),
Schlichenmaier, R. \& Sigwarth, M. (KIS).

\end{document}